\documentclass[journal]{IEEEtran}
\usepackage{blindtext}
\usepackage{graphicx}
\usepackage{caption}
\usepackage{amsmath}
\usepackage[linesnumbered,ruled]{algorithm2e}
\usepackage{amsmath}
\usepackage{algorithmic}
\usepackage{filecontents,lipsum}
\usepackage[noadjust]{cite}
\usepackage{mathtools}
\usepackage{filecontents}
\usepackage[noadjust]{cite}
\usepackage{amsmath}
\usepackage[colorlinks=true,linkcolor=blue,urlcolor=black,bookmarksopen=true]{hyperref}
\usepackage{multicol}
\usepackage{lipsum}
\usepackage{array}
\usepackage{subfig}
\usepackage{textcomp}

\usepackage{atbegshi}
\AtBeginDocument{\AtBeginShipoutNext{\AtBeginShipoutDiscard}}

\begin{document}
%
\title{DeepPos: Deep Supervised Autoencoder\protect\\ Network for CSI Based Indoor Localization}

\author{\IEEEauthorblockN{\normalsize
		Peyman Yazdanian and Vahid Pourahmadi\\}
	\IEEEauthorblockA{\small Electrical Engineering Department, Amirkabir University of Technology, Tehran, Iran}}
	\thanks{Corresponding Author: v.pourahmadi@aut.ac.ir.}


\maketitle

\begin{abstract}

The widespread of mobile devices facilitated emergence of many new applications and services. Among them are location based services (LBS) that provide services based on user's location. Several techniques have been presented to enable LBS even in indoor environments where Global Positioning System (GPS) has low localization
accuracy. These methods use some environment measurements (like Channel State Information (CSI) or Received Signal Strength (RSS)) for user localization. In this paper, we will use CSI and a novel deep learning algorithm to design a robust and efficient system for indoor localization. More precisely, we use supervised autoencoder (SAE) to model the environment using the data collected during the training phase. Then, during the testing phase, we use the trained model and estimate the coordinates of the unknown point by checking different possible labels. Unlike the previous fingerprinting approaches, in this work we do not store the {CSI/RSS} of fingerprints and instead we model the environment only with a single SAE. The performance of the proposed scheme is then evaluated in two indoor environments and compared with that of similar approaches.

\end{abstract}

\begin{IEEEkeywords}
Indoor Localization, Deep Supervised Auto Encoder, Received Signal Strength,
Channel State Information
\end{IEEEkeywords}

\IEEEpeerreviewmaketitle
\section{Introduction}
Wireless technology today has a much wider meaning than the past. Apart from its main application, sending and receiving information, it can be used to facilitate emergence of new applications in various fields \cite{6289200}. In closed environments where GPS satellites are not available, several wireless approaches have been proposed for tasks such as user indoor positioning and navigation \cite{6289200}.

In wireless communications, channel state information (CSI) describes how a signal propagates from a transmitter to a receiver and represents the combined effect of environment such as scattering, fading, and power decay with distance.

Most previous WiFi based localization approaches, mainly used Received Signal Strength (RSS), a parameter that represents the signal strength received from the Access Point (AP). Although collecting RSS and implementing algorithms based on it is very easy due to its availability and low hardware requirements, unfortunately due to multi-path effects and shadowing, it is not a stable measurement for localization, and its  values may change over time irregularly. In addition, RSS values only provide coarse channel information and does not exploit the information existing in many subcarriers of the {orthogonal frequency-devision multiplexing} (OFDM) system that have been used as transmission technology of recent wireless standards \cite{7438932}. 

As an alternative we can use CSI, which is more robust and due to the description of wireless channel in all of its subcarriers, contains more fine-grained characteristics \cite{7438932}. In addition, the CSI values are different for each transmitting-receiving antenna pair, so in cases that the AP or user has multiple antennas, we will have more that one set of CSI measurements at each location which can be useful for tasks such as indoor positioning \cite{7417517}.

Indoor localization approaches are mainly categorized in four major methods which will be explained in next section in detail, but among them fingerprinting based localization schemes is of great popularity. Fingerprinting based approaches usually have two phases \cite{832252}: a) Off-line phase or training stage, which try to construct a database/model with survey data collected at the environment that the localization wanted to be performed afterwards. In this stage, machine learning methods can be applied to determine a compact model for the system instead of storing all the measurements. Several algorithms including recent deep learning method can be used in this stage. b) On-line phase or testing stage, where real time data collected by a user and transmitted to a server. The server then somehow look into all training data and find the most closely matched one(s) and using the selected point(s) suggests an estimation for the coordinates of the user.

Another CSI-based fingerprinting approach is DeepFi \cite{7438932}, which proposes to use CSI measurements for each of the training points and model their statistics with a Deep Belief Network (DBN). In other words, in this method, instead of saving the raw CSI measurements at each training point, it stores the weights of its corresponding DBN. In on-line stage, CSI data of the test point collected and its statistics is compared with all sample points using the fingerprints of the trained models, and the closest matches (with best input CSI reconstruction) are reported as the candidates for target location, and final coordinates is obtained with a probabilistic method. Despite high accuracy, due to the necessity of saving and running all the models/networks, the complexity of the method in both off-line and on-line phases is high and it may be not efficient (specially in environments with large number of training points).

In this paper, we present DeepPos, an indoor localization scheme which is mainly categorized as a fingerprinting scheme, but unlike common other approaches in this group, we do not store fingerprint information for each training points in our database. Instead DeepPos constructs and keeps only one model that represents the whole environment. It is probably the main distinctive feature that separates the proposed scheme from previously known localization methods. To build such model we have used supervised autoencoder (SAE) architecture that has been recently suggested in machine learning community. DeepPos works based on CSI information and it mainly has two stages:

More specifically, in the training phase, we train a supervised autoencoder (SAE) using the raw CSI data that has been collected at a few sample points (SPs) from the environment . Supervised training happens by providing the SP’s ID as the input of the middle layer of the SAE, (more on that in the following sections). After training, the resulted parameters of the network (e.g. its weights and biases) are stored as the environment model.

Afterwards, in the on-line (testing) phase, raw CSI information from the test point is collected and fed as the input of the SAE network generated in the previous stage. 
In order to estimate the coordinates of the unknown point, the SAE tries reconstructing the input CSI by changing its assumption on the point's possible label and calculate the square error between the reconstructed CSI and the input raw CSI for each label. The resulted values are then used in final coordinate estimation of the test point.

The rest of this paper is organized as follows. In Section II, we review the related works in the field of indoor localization. This section is followed by the background information and the preliminaries of our work. Section IV presents the network structure of the DeepPos system. Implementation of DeepPos and experimental evaluations are presented in Section V. Finally, section VI concludes the paper.

\section{Related Work}					
There is a long history of studies in the field of indoor localization and generally they can grouped into four categories based on Ranging, angle of arrival (AoA), time of arrival (ToA) and fingerprints schemes.

\subsection{Ranging-based localization}
Power Ranging-based approaches use at least three access points and the log-distance path loss (LDPL) model and try to estimate location of a node based on received signal strength (RSS) \cite{Lim}. Different approaches have been considered to improve the accuracy of RSS-based methods, mainly by training the LPDL model's parameters \cite{Lim}. EZ \cite{Chintalapudi} presents a no pre-deployment localization schema and uses genetic algorithm to solve RSS-distance identities \cite{7438932}. In another group, range estimation has been carried out using channel state information (CSI) instead of RSS. For instance, FILA \cite{6195606} first uses CSI in different time slots to train the parameters of LPDL model and make it more dynamic. This work also presents some ideas to find the relation between the effective CSI (defined in \cite{6195606}) and distances between node and AP. The trained model is then used for user location estimation.


\subsection{AoA-based localization}
By emergence of WiFi APs with multiple antennas (to support MIMO communications), antenna array based localization techniques, gained lots of interest as well \cite{kotaru2015spotfi}. Indoor localization based on angle-of-arrival (AOA), utilizes multiple antennas to calculate the AoAs of the multipath signals received at each AP. These methods first find the AoA of the direct path to the target, and then apply triangulation to localize the object \cite{kotaru2015spotfi,article2,article3,article4,article5,Sen2013AvoidingMT,paper6}. The main challenge of such techniques is how to improve the resolution of the antenna array \cite{7438932}. SpotFi technique \cite{kotaru2015spotfi}, incorporates super-resolution algorithms that can accurately compute the AoAs of multipath components even when the AP has only three antennas. In this technique there are several APs in the environment, and for each AP, it first estimates the AoAs of all multi-paths in the environment and then finds the AoA of direct path to the target. SpotFi then localizes unknown point using the data it gathered from all APs.
\subsection{Fingerprinting-based Localization}
Fingerprinting-based approaches usually require a training phase to survey the floor plan and a testing phase to search for the best matched fingerprints for location estimation \cite{7438932}. These fingerprints can be built by WiFi \cite{Youssef05thehorus}, RFID \cite{1192765} and Bluetooth technologies. Among different methods and due to availability of WiFi APs in most environments, WiFi based schemes are more common than others. Most previous WiFi based works was mostly done with RSSI. RADAR \cite{832252} is first deterministic work based on WiFi which tries to build fingerprints of RSS using one or more APs and applies K-Nearest Neighborhood (KNN) for position estimation \cite{7438932}. Horus \cite{Youssef05thehorus} is another RSS based scheme, where the RSS from an AP is modeled as a random variable over time and space \cite{7438932}. It identifies different causes for the wireless channel variations and uses  probabilistic techniques to achieve its high accuracy.

RSS based fingerprinting has two main shortcomings \cite{7438932}: First, Due to the multipath effects and shadowing, RSSI is not even fixed for a single point, and they change over time. Second, its values only provide coarse channel information. Instead of RSS, channel impulse response of WiFi is considered as a location-related parameter and contains rich information of wireless channels \cite{7417517}. Similar to ranging based methods, there have been recent studies on application of CSI information (instead of RSS) in fingerprinting application too. For instance, FIFS \cite{6289200}, presents an idea to find the relation between the effective (defined in \cite{6195606}) and the distances of the node and the AP. It also proposes a probabilistic algorithm based on coherence bandwidth which enhances the accuracy of mapping the test point to the fingerprints. The shortcoming of this model is that it needs a large amount of calibration points to build the fingerprint database \cite{7438932}.

DeepFi \cite{7438932} is another powerful CSI-based fingerprinting system. This method first collects CSI at some locations as training spots and for each of the them, constructs a deep neural network to model the statistics of that point and store all the models of all training spots as its database. For localization, CSI data is collected at a test point and its statistics is compared with the models which have been constructed in the training phase for each point and the closest matches are then selected as the candidates for the test point. Although it has high accuracy, the issue of this scheme is its very high complexity in both off-line and on-line phases to construct and check all models. It is also necessary to save parameters of all trained networks for each sample point, which may be not that efficient, specially when the number of  training points increases (to have higher localization accuracy).

\section{Background Theory And Preliminaries}
Here we first bring an overview of modulation technique in WLAN and its channel status information (CSI). Then we bring a section that briefly reviews basics of deep neural networks specially AEs.

\subsection{Channel State Information in OFDM}
Orthogonal Frequency Division Multiplexing (OFDM) technique has been widely used in different technologies like IEEE802.11a/g/n, WiMAX, LTE standards to access the high data rate. In this technique, first, in the transmitter, the input data stream is broken into several streams each of which transmitted over non-overlapped subcarriers after they modulated with a certain modulation scheme. The data then converted back to the time domain by an inverse Fast Fourier Transform (IFFT). The data is then serialized and converted to analog signal and sent through the channel. At the receiver, after returning to the digital domain by Fast Fourier Transform (FFT), the signal is equalized (the effect of channel is removed) in the frequency domain. Demodulation and decoding are next steps in the receiver chain \cite{6289200}.

In WiFi Network Interface cards (NIC) usually all the above steps are happening inside of the chipset and are not accessible from outside. Thanks to an open source Linux kernel \cite{Halperin_csitool}, and with few NICs, such as Intel IWL 5300, it is now possible to access the CSI from outside of the NIC and record the channel variation experienced during transmission of one OFDM sample over all subcarriers. Without CSI. As for notation, let $\mathbf{x}_i$ and $\mathbf{y}_i$ denote the transmitted and received signal at each subcarrier. We have 
\begin{equation}
\mathbf{y}_i= \mathbf{H}_i\mathbf{x}_i + \mathbf{n}_i
\end{equation}
where $i$ denote the number of subcarrier, $\mathbf{n}_i$ is the additive white Gaussian noise and $\mathbf{H}_i$ represents the channel gain at the $i^{th}$ subcarrier. 

The WiFi channel at the 2.4 GHz band can be considered as collection of 56 narrowband  sub-channels. From these 56 subcarriers, the Intel WiFi Link 5300 NIC driver implementation is able to report back the CSI of 30 subcarriers. Its reported channel frequency response is a complex value fro each subcarrier, which is the real and imaginary part of the $\mathbf{H}_i$:\\

\begin{equation}
\mathbf{CSI}_i = \mathbf{H}_i = \mbox{real}(\mathbf{H}_i)+\mathbf{j}.\mbox{imag}(\mathbf{H}_i) 
\end{equation}
In this paper, the proposed DeepPos framework is based on the CSI of 30 subcarriers,
which can reveal much richer channel properties than RSSI.

\subsection{Deep Learning Approaches} 

Application of machine Learning techniques in different problems is growing very fast due to their vast capability for solving real problems. Availability of massive amounts of computational power and emergence of some methods for solving the vanishing gradient problem of neural networks, make it practical to use large neural networks, and in particular deep networks \cite{Deep}. Autoencoder (AE) is an important class of deep learning networks which usually are categorized as unsupervised learning algorithm and its main objective is to train network such that the target values (at the output nodes) be equal to the inputs.

\begin{figure}
	\includegraphics[width=3.7in,height=2.5in,clip,keepaspectratio]{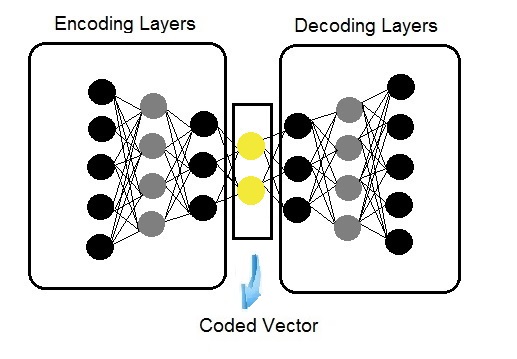}
	\caption{Structure of a Deep AE}
	\label{aut}
\end{figure}

A deep AE is composed of two symmetrical deep-belief networks that typically have four or five fully connected layers representing the encoding half of the net, and second set of four or five layers that make up the decoding half. Figure \ref{aut} shows a simplified schema of a deep AE’s structure. As an unsupervised learning algorithm, AE determines the weights of the network based on back propagation such that it minimizes the loss function of $L=\|\mathbf{y}-\mathbf{x}\|^2 $, where $\mathbf{y}$ and $\mathbf{x}$ are the output and input of the AE respectively. 

An AE in essence is trying to learn an approximation to the identity function \(h_{W,b(x)}(x)\), so as to output \(y\) is equal to input \(x\) \cite{Autoencoders}. The hidden layers are Restricted Boltzmann Machines (RBM), which is the building blocks of deep-belief networks \cite{RBM}. The middle layer of the AE (which usually has the fewest number of neurons) generally can be considered as the low order representation of the input vector.

\section{DeepPos System}

DeepPos has two major phases, off-line training stage, and on-line location estimation
that we will discuss them in the following subsections. First, in training phase, we model the environment via a deep neural network in the form of SAE. Then, in on-line stage, using this trained model, we select a few candidate points which will be then used to estimate the actual coordinates of the test point.

\subsection{DeepPos Architecture}  
Supervised autoencoder (SAE) which is a new architecture in deep learning community,
is the core of our scheme. As mentioned in previous section, the main distinctive feature of DeepPos is training only one model for the whole environment and so there is no need to store many models or fingerprints information for each SP. 

\begin{figure}
	\includegraphics[width=3.6in,height=5in,clip,keepaspectratio]{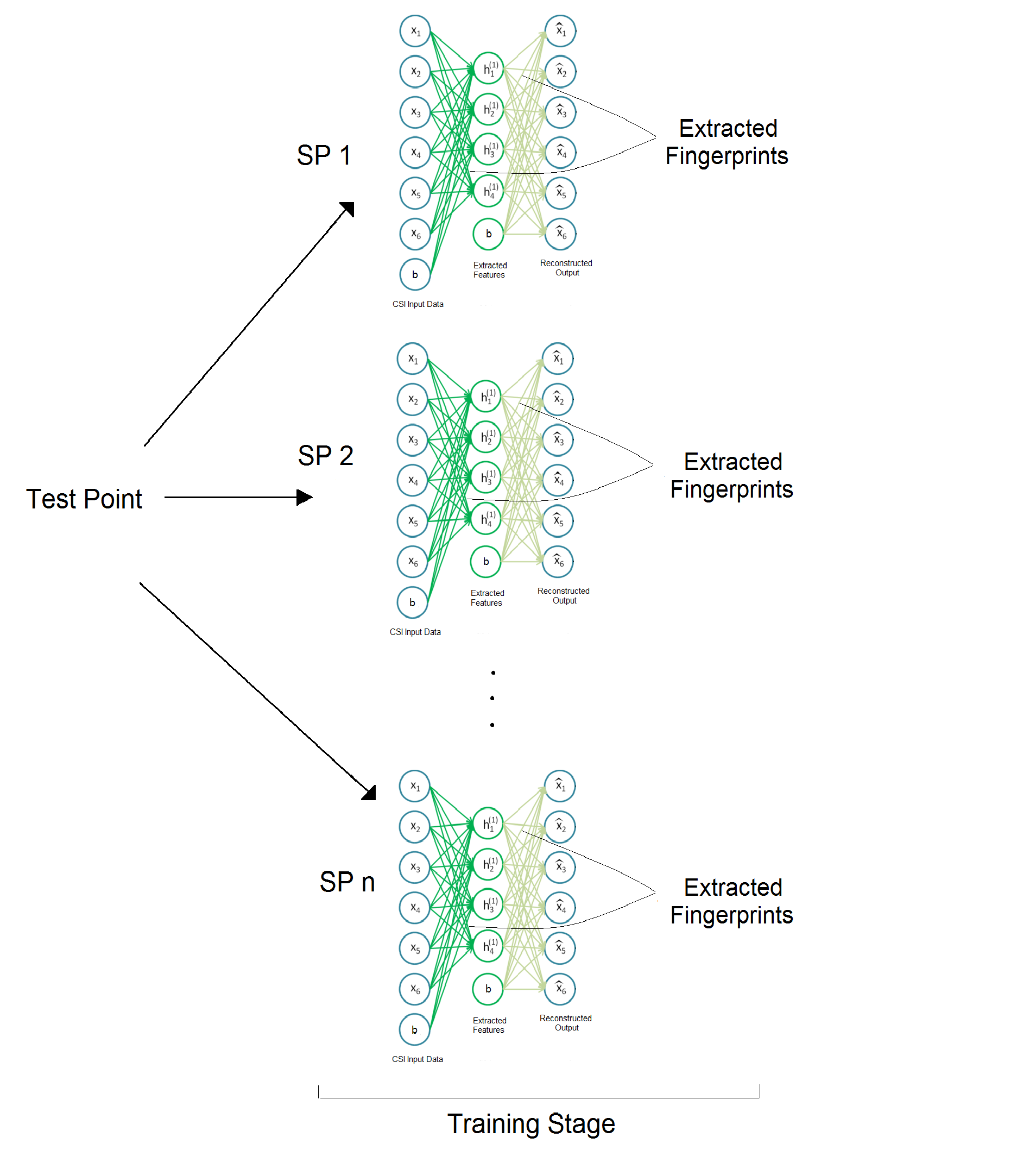}
	\caption{DeepFi Main Architecture}
	\label{deep_fi_Architecture}
\end{figure}

Additionally, in previous deep learning models like DeepFi \cite{7438932}, the individual Deep Belief Networks (constructed for each SP) is designed in an unsupervised manner (it does not have any information about the location of the collected CSI). Figure \ref{deep_fi_Architecture} shows the overall architecture of the DeepFi system.  In contrast, DeepPos transforming raw input CSI to a lower dimension in a supervised manner, i.e. it tries to extract features that not only they are good for input reconstruction (like autoencoder) but also these features should keep information related to the location of the collected CSIs. Due to such structure (that has both lower dimension and the location information), it is possible to have only one single model for the whole environment. 

\begin{figure}
	\vspace{0.6cm}
	\includegraphics[width=3.6in,height=5in,clip,keepaspectratio]{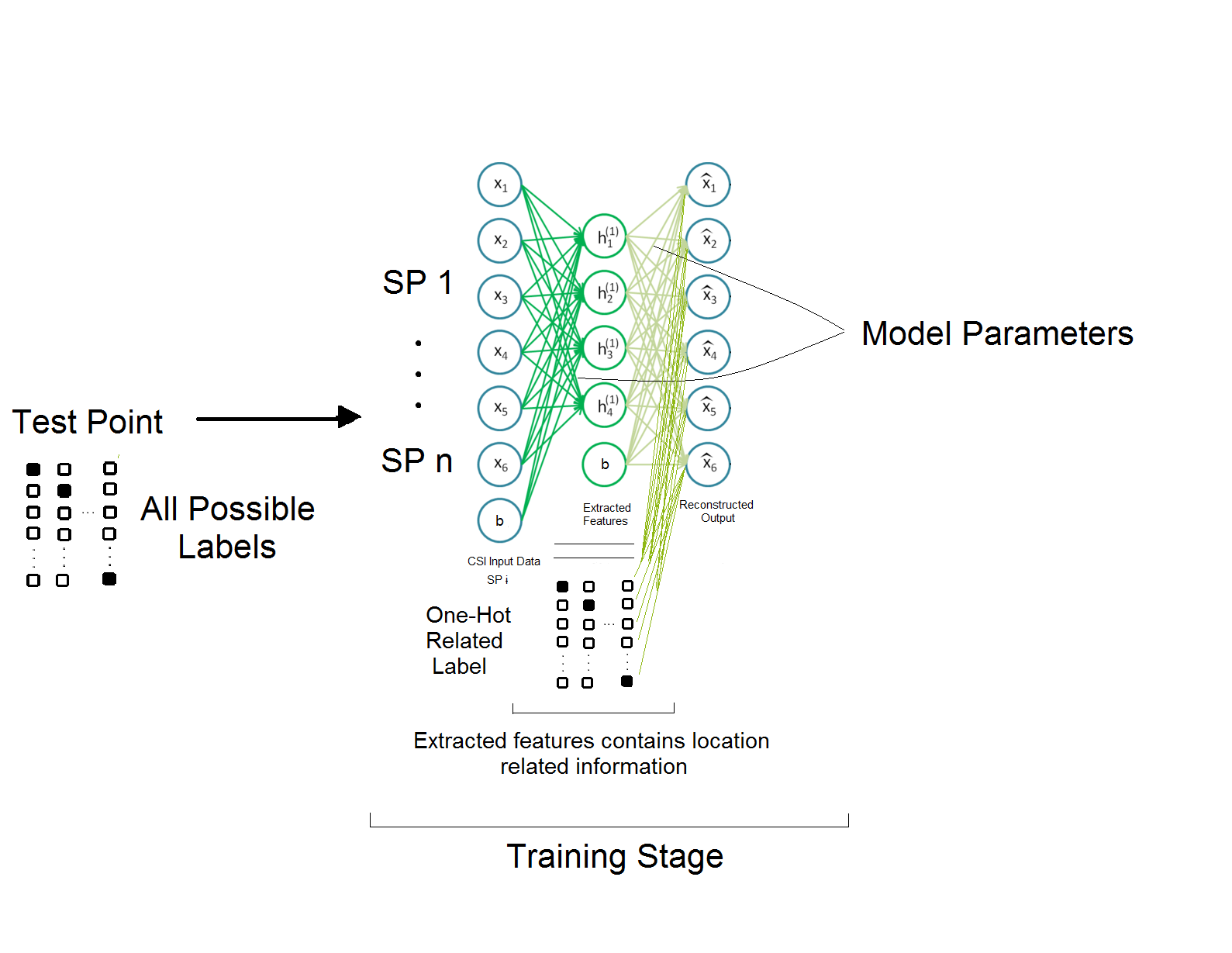}
	\caption{DeepPos Main Architecture}
	\label{dpos_Architecture}
\end{figure}

Figure \ref{dpos_Architecture} depicts the overall architecture of DeepPos. As can be seen, DeepPos mainly have 3 parts: first, the encoding layers acts as a feature extractor. Then, the extracted features are concatenated with an one-hot vector representing the label for the training data. Finally, the combined vector is used as the input of the decoding layers which aims to reconstruct the original CSI vector. Detail explanation of each part is presented in section \ref{ssec:train}


\subsection{Training Stage With Deep Learning}  \label{ssec:train} 
In this phase, raw CSI data are collected at some SPs of the environment, each contains CSI measurements of  the 30 OFDM subcarriers of 3 antennas. The diversity of data makes it possible for the system to learn more about the environment and to produce a more comprehensive model. In this work we only consider the amplitude responses of collected CSIs and the values are normalized between (0,1). 

\begin{figure*}
	\centering
	\includegraphics[width=0.8\textwidth]{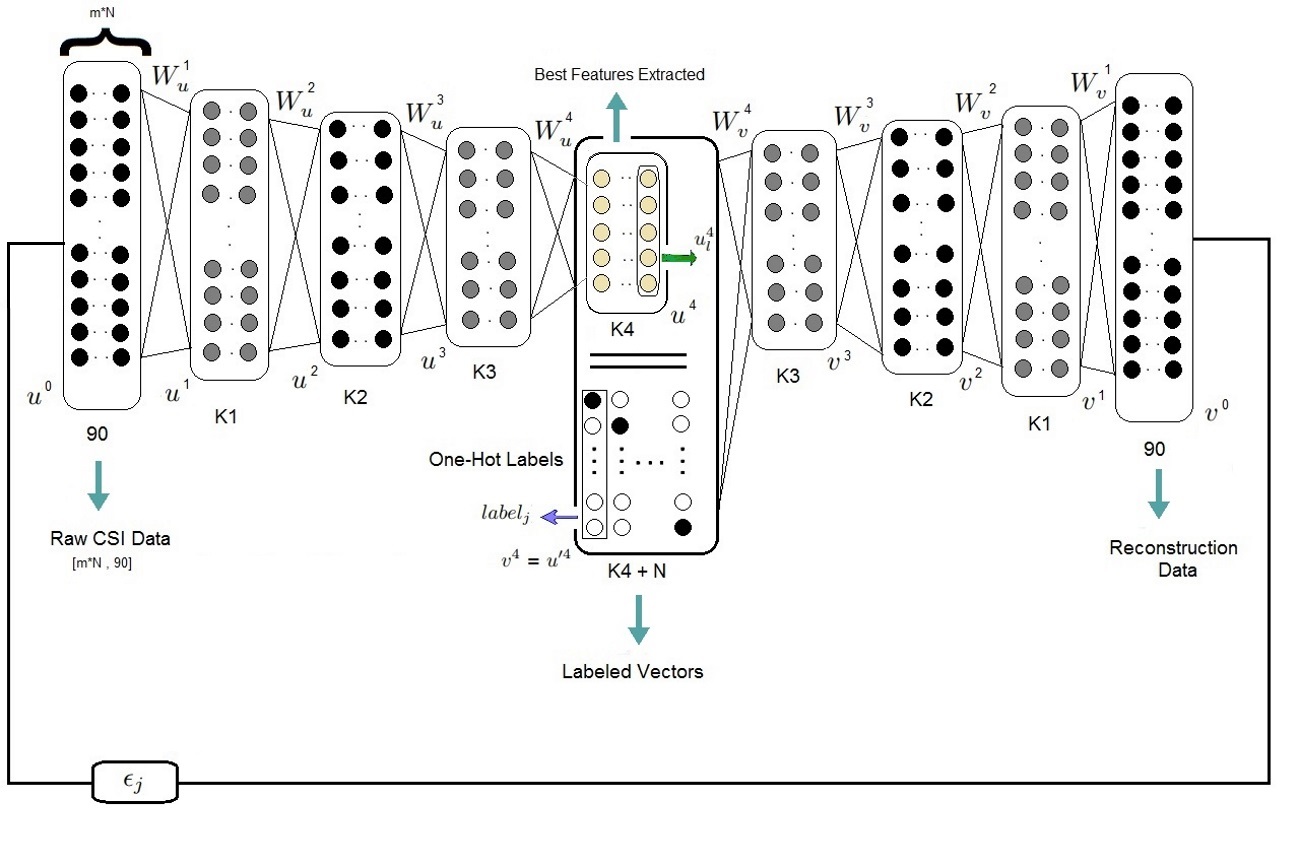}
	\caption{DeepPos Training Stage Schema}
	\label{off}
\end{figure*}

The collected data is used to train a a deep supervised autoencoder which aims to reconstruct the input CSI values. Figure \ref{off} shows overall architecture of DeepPos training stage. A deep SAE network with four hidden layers is used to train and fit a model. In order to reduce the dimension of raw CSI data, we assume that the number of neurons in hidden layers are gradually decreasing. We denote the number of neurons in the first to the fourth layer by \(K_1 , K_2 , K_3 , K_4 \)  where \(K_1 > K_2 > K_3 > K_4 \). 

The pseudocode for the training stage is presented as in algorithm \ref{alg1}.
To model the scenario with only one deep neural network, in each training iteration, after encoding of input data from each of the CSI measurements (lines 6-9), we append its corresponding label as a one-hot vector to the output of the middle layer of the network (lines 10-11). The resulted vector is then fed to the decoding layers to reconstruct the input CSI data (lines 12-17). The aim is to find network weights minimizing the reconstruction error. 

We define \( u^i_l\) as hidden variable in encoding layer \(i\), for \(i\) = 1,2,3,4, and let \(\mathbf{d}^j_l\) denote the $l^{th}$ measurement vector of the $j^{th}$ SP. vector \(\mathbf{d}^j_l\) is of size 90, corresponding to 30 subcarriers measurement at 3 antennas, collected at a training spot. At each step, $u^0_l$ gets in one of the measured vectors, \(\mathbf{d}^j_l\), as the network input. Furthermore, as Sigmoid function is used as the activation function of each layer and so we have:
\begin{multline}
u^i_l = \mbox{sigmoid} (W_{u}^i. u^{i-1}_l +  b_{u}^i) \hspace*{1.5ex} i = 1,2,3,4
\label{eq1}
\end{multline}
\(W_{u}^i\) and \( b_{u}^i\) are the weights and biases matrix for encoding layer \(i\) respectively.

After the encoding operation, in layer 4, we append \( u^4_l\) with appropriate label as an one-hot vector. For example, if the input CSI was the $l^{th}$ CSI measurement of the $j^{th}$ SP (from total of $N$ SPs), we append an one-hot vector corresponding to $j$ (A vector of size $N$ with all zeros except at location $j$ which has one). 
So \( u'^4_l\) can be defined as:
\begin{equation}
u'^{4}_l = (u^{4}_l,label_{j})\hspace*{3ex}
\label{eq2}
\end{equation}
Where \( u^{4}_l\) is the encoded representation of \(\mathbf{d}^j_l\) and \(label_{j}\) is the label one-hot vector for that packet. Note that the main difference between the common autoencoder and the supervised autoencoder is in addition of the labels at the middle layer of the network.  

Finally, for the decoding stage we have:
\begin{multline}
\hspace*{5ex}v^4 = u'^4; \\\
v^i = sigmoid (W_{v}^{i+1}. v^{i+1} + b_{v}^{i+1})\hspace*{1.5ex} i = 3,2,1,0
\label{eq3}
\end{multline}
Where \(W_{v}^i\) and \( b_{v}^i\) are the weights and bias matrices for decoding layer \(i\), respectively. Like common autoencoders, the goal is that the output of the network (\( v^0\)) be the best possible reconstruction of the input CSI vector. So, at it each epoc, the network weights are updated such that they minimize the cost function defined as:\protect\\
\begin{equation}
\epsilon= \sum\limits_{l=1}^{N} \sum\limits_{j=1}^{m}  \| v^{0}_l - \mathbf{d}^j_l\|^2 
\end{equation}
Where \( v^0_l\) is the output of the network for the input of $\mathbf{d}^j_l$.
In this work we have used RMSprop optimizer as the deep network optimization technique. The resulted network guaranties that for each input CSI measurement of a SP, it gets the lowest reconstruction error for a label that matches the point's ID.

\begin{algorithm}                           
\SetKwInOut{Input}{Input}
\SetKwInOut{Output}{Output}

\Input{$m$ measurements [$\mathbf{d}$] each with 90 CSI values collected and stored for each of the $N$ training spots}
\Output{Model parameters (Trained Network's weights and biases)}

Initialize weights of all layers \( W_{u}^i\), \( W_{v}^i\), \( b_{u}^i\), \( b_{v}^i\) = randn // randn is standard Gaussian distribution function\protect\\
\For{$k=1$ : max-epoch}
{	
	\For{$j=1$ : $N$}
	{
		\For{$l=1$ : $m$}
		{	
			\( u^0_l\) = the $l^{th}$ measurment vector of the $j^{th}$ SP \(\mathbf{d}^j_l\) ;\protect\\	
			// encoding\protect\\ 
			\For{$i=1$ : $4$}{
				\(u^{i}_l\) = sigmoid (\(W_{u}^i\). \(u^{i-1}_l\) + \( b_{u}^i\))\protect\\
		}
		// append appropriate label for each input vector \protect\\
	
		\(u'^{4}_l\) = (\(u^{4}_l\),\(label_{j}\));\protect\\
	
		// decoding\protect\\
			\(v^{4}_l\) = 	\(u'^{4}_l\);\protect\\
		\For{$i=3$ : $0$}{
			
			\(v^{i}_l\) = sigmoid (\(W_{v}^{i+1}\). \(v^{i+1}_l\)+ \( b_{v}^{i+1}\))\protect\\	
		}
		Obtain the error between input data \(\mathbf{d}^j_l\) and reconstructed
		data \( v^0_l\);\protect\\
	}
}

	//fine-tuning;\protect\\

	Update network weight and bias based on the resulted errors with
	back-propagation with Adam optimaizer;
}

\caption{Off-line Training Stage}
\label{alg1}
\end{algorithm}

\subsection{On-line Location Estimation With Consider KNN Points} 
The procedure of the DeepPos on-line stage is presented in algorithm \ref{alg2}. First, DeepPos takes a few CSI measurements at the test point and feed them to the trained SAE model. Note that for SAE to reconstruct the CSI, it needs to have a label (in the form of one-hot vector) input at its middle layer, but for the test location we do not have that label (since we do not know the location of the test point). The way that DeepPos proceed is to loop over all possibilities of one-hot vectors (corresponding to different SPs) and for each case, it calculates the mean square error (MSE) of the reconstructed CSI for the collected measurements. The location of the test point is then calculated using the weighted sum of the SP's coordinates corresponding to the lowest MSEs.

\begin{figure}
\hspace*{10ex}\includegraphics[width=4in,clip,keepaspectratio]{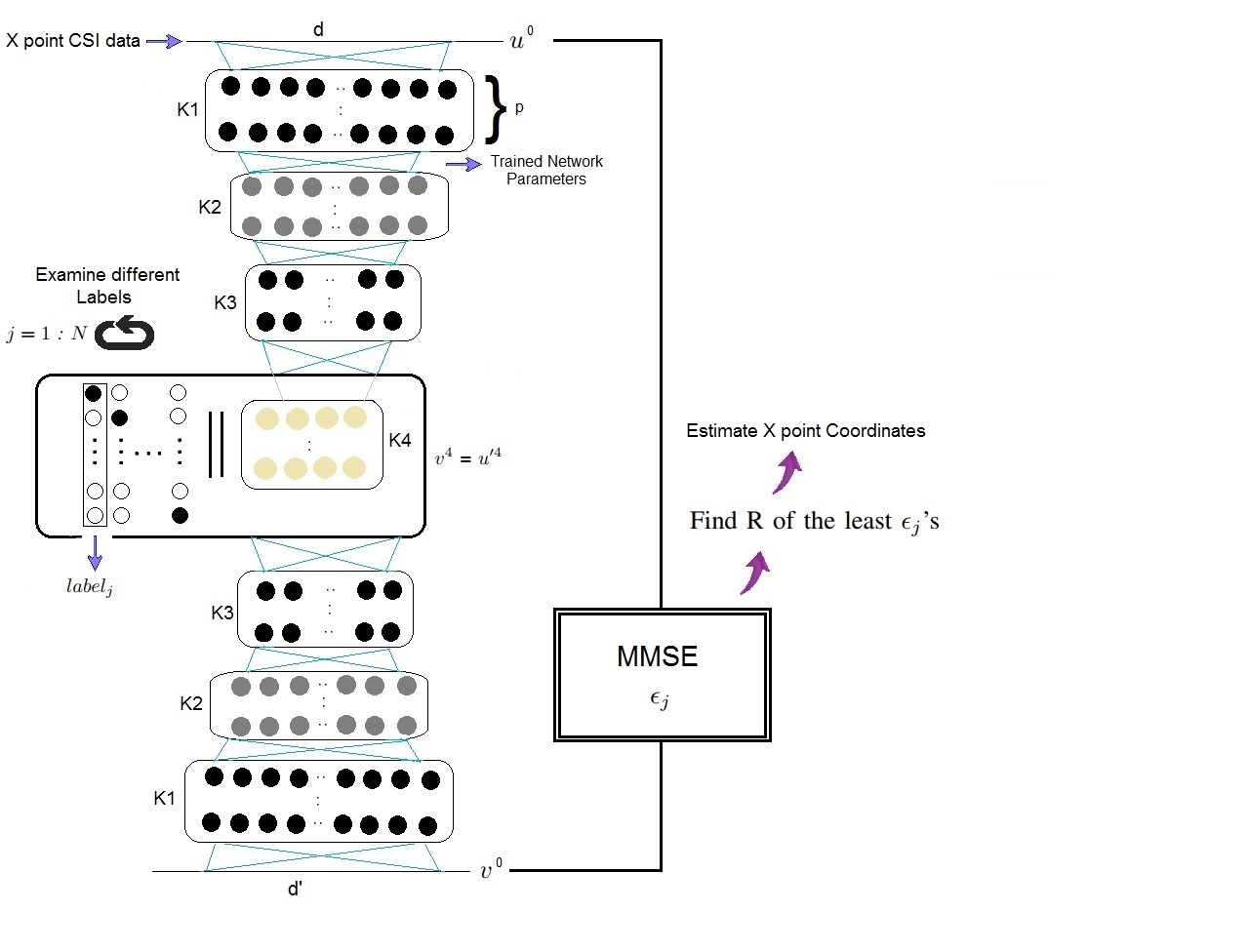}
\caption{DeepPos On-line Location Estimation Stage Schema}
\label{fig3}
\end{figure}

In algorithm \ref{alg2}, the encoding layers construct the coded features of the unknown point's CSI raw data (lines 1-7). Then, for each of the $N$ SPs, its corresponding one-hot vector is appended to this coded features. The decoding layers generate the reconstructed vector of the input (lines 7-16), and the mean square of the reconstruction error is evaluated. To have a more robust estimation, the MSE calculation is performed for $p$ CSI measurements (at the test point), and averaged over all of them. Labels that lead to a lower reconstruction MSEs are the candidate points from the network's perspective for the test spot (lines 17-21). Finally, the test point's coordinates are evaluated as the weighted mean of these candidate points (line 23). Figure \ref{fig3} shows overall schema of DeepPos on-line stage.\protect\\

\raggedbottom
\begin{algorithm}
\SetKwInOut{Input}{Input}
\SetKwInOut{Output}{Output}

\Input{$p$ measurements $\mathbf{d}$ each with 90 CSI values collected at the test point,\\
	Trained SAE model}
\Output{Estimated test point's coordinates;}

// encoding\protect\\
\For{$l=1$ : $p$}{
	\( u^0_l\) = $l^{th}$ measurment vector of the test point \(\mathbf{d}_l\) ;\protect\\				
	\For{$i=1$ : $4$}{
		
		\( u^i_l\) = sigmoid (\(W_{u}^i\). \(u^{i-1}_l\) + \( b_{u}^i\))\protect\\	
	}
}
// append different labels to low order representation vector (\( u^4_l\)), and find the reconstruction of the input data\protect\\
\For{$j=1$ : $N$}
{
	
	\For{$l=1$ : $p$}{

		\(u'^{4}_l\) = (\(u^{4}_l\),\(label_{j}\));\protect\\
		// decoding\protect\\
		\( v^4_l\) = \( u'^4_l\);\protect\\
		\For{$i=3$ : $0$}{
			
			\( v^i_l\) = sigmoid (\(W_{v}^{i+1}\). \(v^{i+1}_l\) + \( b_{v}^{i+1}\))\protect\\	
		}
		\(err_{l} =  \| v_{l}^{0} - d_{l} \|^2\)\protect\\
	}
	
	$\epsilon_{j} = \frac{1}{p} \sum_{l=1}^{p} err_{l}$\protect\\
	
}	

// Find R of the least $\epsilon_{j}$s \protect\\
// Estimation coordinates of unknown point\protect\\
\((x\hat{},y\hat{})\) = $\sum_{r=1}^{R}$\(\frac{1}{\epsilon_{j}}\) \((x\hat{}_{r},y\hat{}_{r})\) ;\protect\\
\vskip 0.2in

\caption{On-line Location Estimation Stage}
\label{alg2}
\end{algorithm}

\section{Experimental Evaluation}
We implemented DeepPos with 2.4 GHz commodity WiFi Intel 5300 NIC and carry out different experiments to validate its performance. We have used a Dell laptop equipped with an Intel 5300 NIC as the access point, and a Huawei handset as the mobile device. We have used the modified Linux driver developed by \cite{Halperin_csitool} to measure and store CSI information into a file. These files are sent to the server (which is a linux machine with 8 GB of memory and Intel\textregistered\ Core\texttrademark\  i5-3210M processor), and the CSI data for each packet from each training points are stored in a PostgreSQL database.
Note that if the training SPs are too sparse, it may result in low localization accuracy in testing phase as there are not enough data to model. On the other hand, if we choose dense training spots, it will cost a lot of efforts on pre-training data collection. Based on our experiments, the distance between two spots is set to about 1m, which can maintain the balance between localization accuracy and pre-processing cost.

\subsection{Experiment Configuration}  
We verify the performance of DeepPos in two indoor scenarios. In each case, we collect data from a few  SPs and at each SP we measure 30 packets continuously each of them contains 90 raw CSI data (30 subcarriers in 3 antennas). All the measurements are collected at a same height, which construct a 2D platform. First one is an empty classroom where there is a LoS path between the AP and mobile device. Second location is a room and corridor around, which is more challenging scenario as there are some points without LoS path. We evaluate overall system performance in both settings by using one leave-out cross-validation test. The experimental results for both scenarios are presented in the following.\protect\\
\raggedbottom
\begin{subsubsection}{Class Room in the University} 
Figure \ref{sn1} shows the layout of our first test scenario. In this 6$\times$7 \( m^2 \) classroom, the AP is located on the table at one end of the room, and all the training and test samples are collected at almost same height. As shown by red dots in figure \ref{sn1}, SP positions are chosen uniformly scattered 
in the room. We collect 30 packets for each of the 19 SP. We choose deep network's parameters as \( K_1 = 50, K_2 = 30, K_ 3 = 20,\) and \(K_4 = 5\).
\begin{figure}
	\hspace*{3ex}\includegraphics[width=3.2in,clip,keepaspectratio]{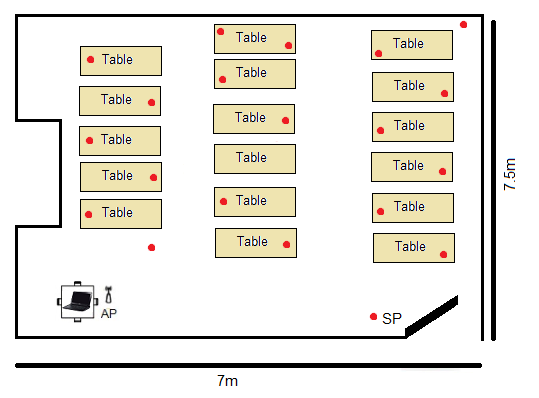}
	\caption{Classroom Layout}
	\label{sn1}
\end{figure}

\begin{figure}
	\hspace*{3ex}\includegraphics[width=3.3in,clip,keepaspectratio]{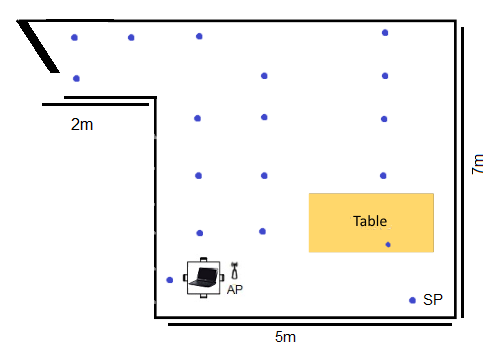}
	\caption{Hall and Corridor Layout}
	\label{sn2}
\end{figure}

\end{subsubsection}
\begin{subsubsection}{Empty Hall and corridor around}
The hall and its entrance corridor is the second scenario in which there is some points are not in the LOS of the AP. The layout of this location is presented in figure \ref{sn2} with dimensions of almost 5$\times$7 \( m^2 \). As shown in figure \ref{sn2}, The layout of this location .The SAE structure is similar to the first scenario and \( K_1 = 50, K_2 = 30, K_ 3 = 10,\) and \(K_4 = 5\).

\end{subsubsection}

\subsection{System Performance With Cross-Validation Test}
To test the overall functionality of the system, leave one out cross-validation (LOOCV) test is applied. Specially when we have limited number of samples, LOOCV often gives us more accurate results compared with simple partitioning of data into train and test sets. More precisely, in LOOCV, at each step, the data of one SP is set aside (for testing) and the SAE is trained with the data of remaining SPs. Then in the on-line stage, we randomly select $p$ measurement data from the test SP and feed that into the the trained model. The model then reports back 2 (or 3) best candidate SPs that has less reconstruction MSE when we use their corresponding one-hot vector in the SAE. The final estimated location of the test point is then reported back as the weighted mean of these candidate spots.

The Euclidean distance between the estimated coordinates and the actual one of the test point is defined as the localization error for that SP.  Denoting the estimated coordinates for the $i^{th}$ SP by \((\hat{x}_i,\hat{y}_i)\) and the actual position by \((x_i,y_i)\) we will have
	$\epsilon_i = \frac{1}{N} \sum_{n=1}^{N} \sqrt{(\hat{x}_i-x_i)^2+ (\hat{y}_i-y_i)^2}$.
The mean of all localization errors for all SPs is what we reported as the performance of the system. 

To see the performance of the DeepPos, we compared its localization error with that of DeepFi \cite{7438932}. As a quick reminder, DeepFi was one of the most recent indoor positioning scheme that uses deep learning method but it constructs a separate deep model for each of the SPs; thus, each model represents one of them and we need to have N deep models. Unlike DeepPos, where we only have a single deep network which represents the whole environment (not a single SP).  

\begin{figure}
	\centering
	\includegraphics[width=3.5in,clip,keepaspectratio]{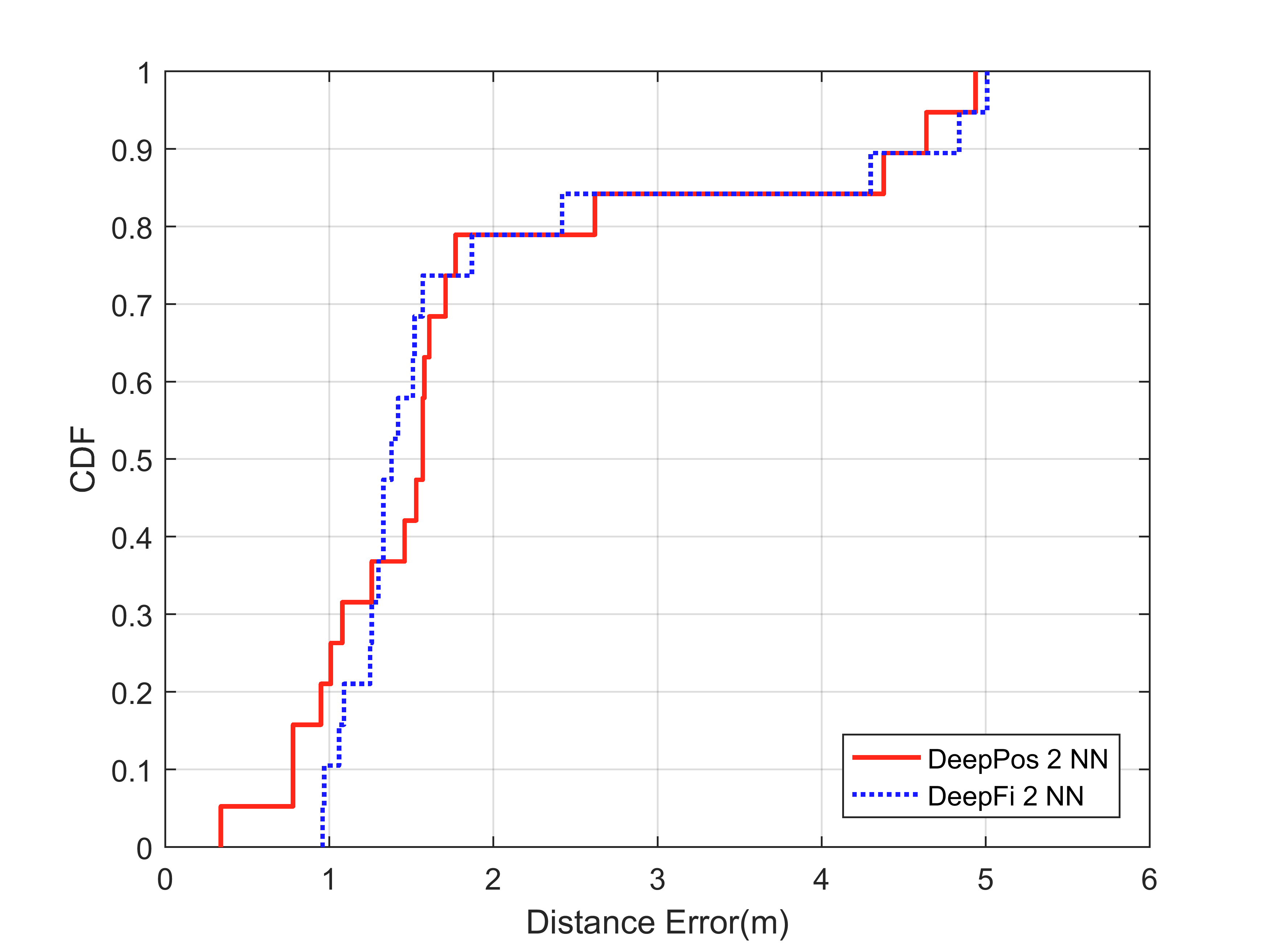}
	\caption{CDF of localization error in Classroom}
	\label{cdf2}
\end{figure}

Figure \ref{cdf2} shows the CDF of localization errors when we use 2NN DeepPos and DeepFi  in the first environment (classroom scenario). DeepPos has mean and median localization error in this case is about 1.87m and 1.57m respectively; while they are 1.91m and 1.38m for DeepFi respectively. In this scenario, DeepPos has 80\% of localization errors under about 2m.

\begin{figure}
	\centering
	\includegraphics[width=3.5in,clip,keepaspectratio]{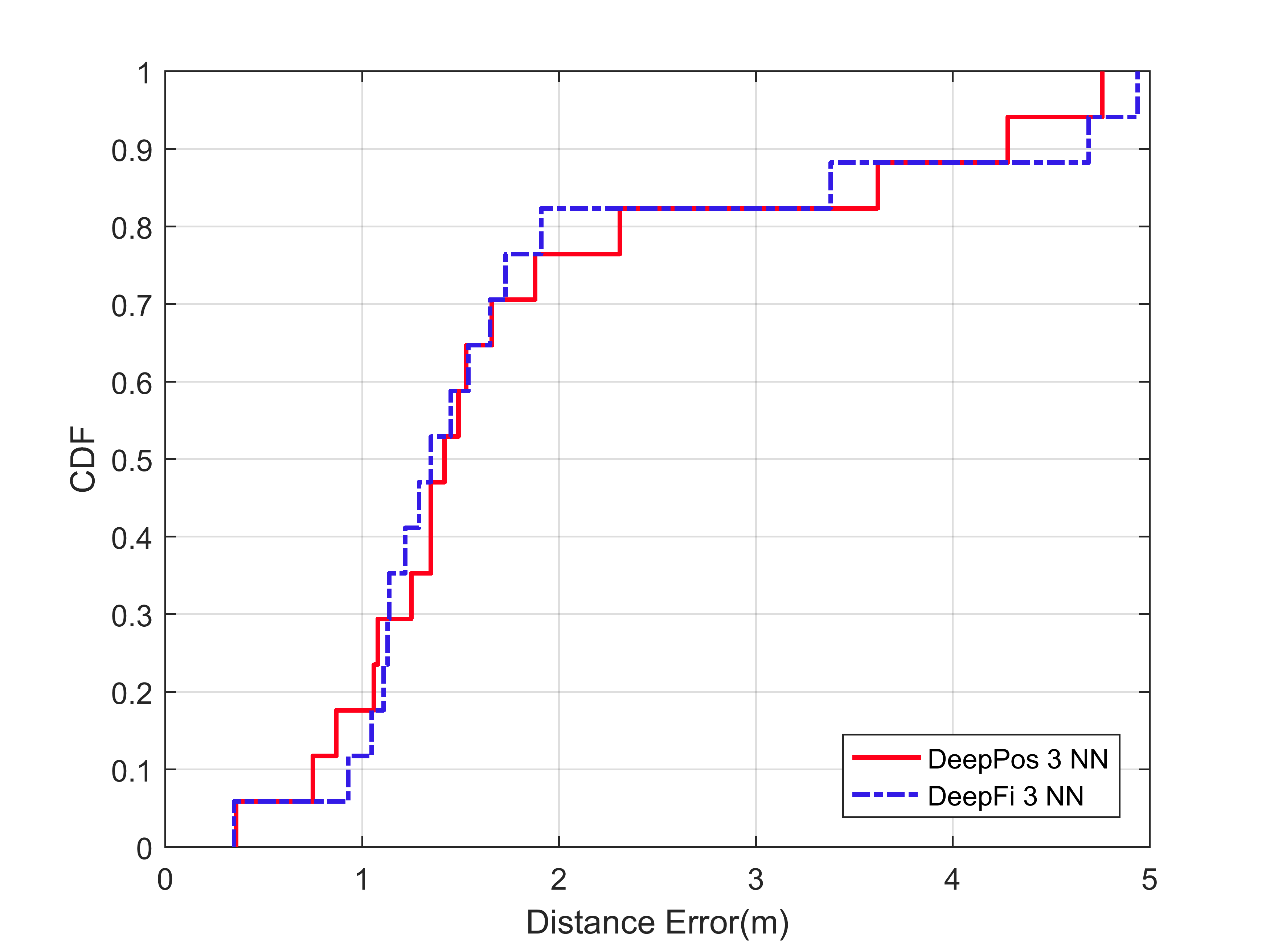}
	\caption{CDF of localization error in Hall and Corridor}
	\label{cdf3}
\end{figure}

Similar experiment has been conducted for evaluating DeepPos performance in the second environment. Figure \ref{cdf3} shows CDF of distance errors in this place for 3NN DeepPos and DeepFi. In this case, DeepPos and DeepFi have the mean error of 1.82 and 1.81, respectively. For DeepPos, 80\% of the localization error is less than 2m.

Table \ref{tab:a} summarizes the overall performance of DeepPos and DeepFi in both scenarios. This results shows that two systems have close localization performance, however, different from DeepFi, the proposed DeepPos system does not require to train a deep neural network for every training spots, which makes the model very complex. Instead, by using a single deep supervised autoencoder, we model the whole environment. Table \ref{tab:a} also presents the overall average times spent for LOOCV test of two systems both exceeded on the same environment and with the same dataset, which strongly shows simplicity of the proposed method compared to DeepFi.

\begin{table} 
	\hspace*{3.5cm}\caption{\\*Compare of Mean, Std. of error and Exe. time (s)\\*}\label{tab:a}
	\begin{center}
		\begin{tabular}{  m{7cm}  }
			\vspace*{0.5ex}
			\hspace*{3cm}  Classroom \\
			
		\end{tabular} 
		\begin{tabular}{ | m{1.2cm} | m{1.8cm}| m{1.5cm} | m{2cm} | } 
			\hline
			Method & Mean error (m) & Std. error (m) & \vspace*{1.5ex} Average Execution Time (s) \\
			\hline
			\vspace*{0.7ex}
			\vspace*{0.7ex}
			DeepPos & 1.872 & 1.331 & 1.1449\\ 
			
			DeepFi & 1.915 & 1.295 & 3.7072\\ 
			\hline
		\end{tabular}
	\end{center}
	
	\vspace*{0.5cm} 
	\begin{center}
		\begin{tabular}{  m{7cm}  }
			\vspace*{0.5ex}
			\hspace*{3cm}  Hall and Corridor \\
		\end{tabular} 
		\begin{tabular}{ | m{1.2cm} | m{1.8cm}| m{1.5cm} | m{2cm} | } 
			\hline
			Method & Mean error (m) & Std. error (m) & \vspace*{1.5ex} Average Execution Time (s) \\
			\hline\vspace*{0.7ex}
			\vspace*{0.7ex}
			DeepPos & 1.824 & 1.240 & 1.053 \\ 
			
			DeepFi & 1.815 & 1.287 & 3.457\\ 
			\hline
		\end{tabular}
	\end{center}
\end{table}


\subsection{Number Of Sample Points And Overhead Analysis} 
In this section, we study the complexity of the localization framework in both off-line and on-line phases. \\

\subsubsection{\textbf{Evaluating Training Overhead}}

In this part, we evaluate the training stage complexity and overhead in terms of total time required to train a model and memory usage as the number of SPs increases.  
To get an average time/RAM usage/accuracy, in each case, we have selected all possible combinations for training SPs and averaged the results over all cases. For example, if we want to report the average training time when we have $k$ SPs, we select all possible combinations of $k$ SPs from all SPs (in that scenario) and report the average training time of all cases.

\begin{figure}[!tbp]
	\centering
	\begin{minipage}[b]{0.23\textwidth}
		\includegraphics[width=\textwidth]{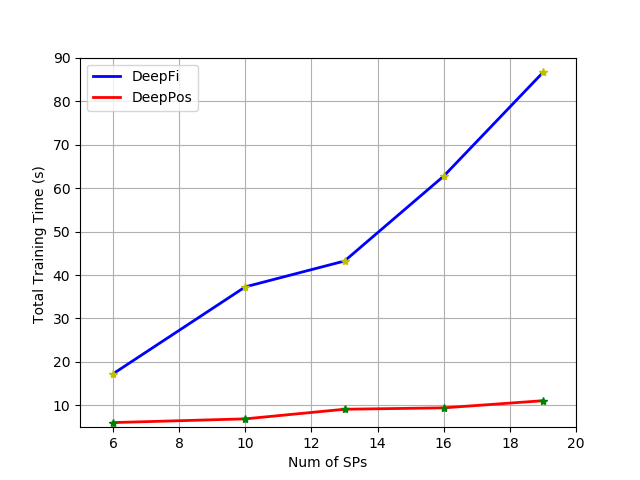}
		\caption*{a) Classroom}
	\end{minipage}
	\hfill
	\begin{minipage}[b]{0.23\textwidth}
		\includegraphics[width=\textwidth]{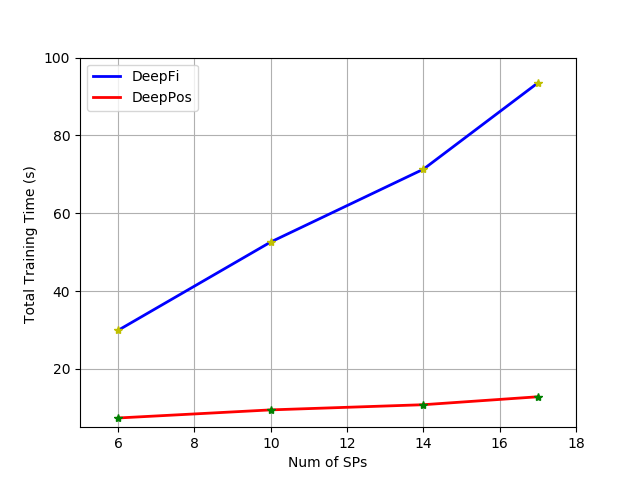}
		\caption*{b) Hall and Corridor}
	\end{minipage}
	\caption{Training stage time spent for sequential test}
	\label{time-test1}
\end{figure}

\begin{figure}[!tbp]
	\centering
	\begin{minipage}[b]{0.23\textwidth}
		\includegraphics[width=\textwidth]{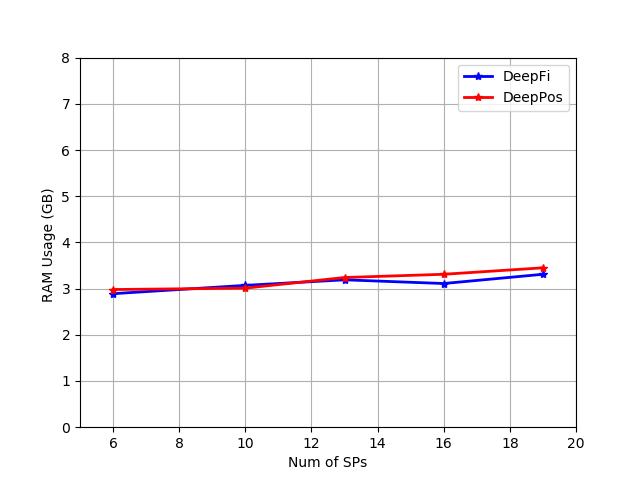}
		\caption*{a) Classroom}
	\end{minipage}
	\hfill
	\begin{minipage}[b]{0.23\textwidth}
		\includegraphics[width=\textwidth]{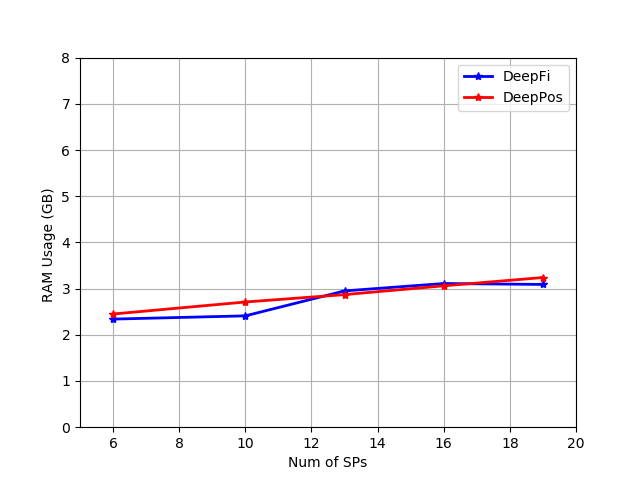}
		\caption*{b) Hall and Corridor}
	\end{minipage}
	\caption{Training stage memory usage for sequential test}
	\label{ram-test1}
\end{figure}

Note that, as DeepFi constructs an individual model for each of the SPs, we can assume that these models are trained sequentially which keep the memory usage of that scheme in a normal range but the training time increases linearly as the number of SPs grows. With this assumption, figures \ref{time-test1} and \ref{ram-test1} shows the results for both DeepPos and DeepFi. As can be seen, DeepPos needs much less training time (specially when number of SPs increased), while they have almost similar memory usage. 

\begin{figure}[!tbp]
	\centering
	\begin{minipage}[b]{0.23\textwidth}
		\includegraphics[width=\textwidth]{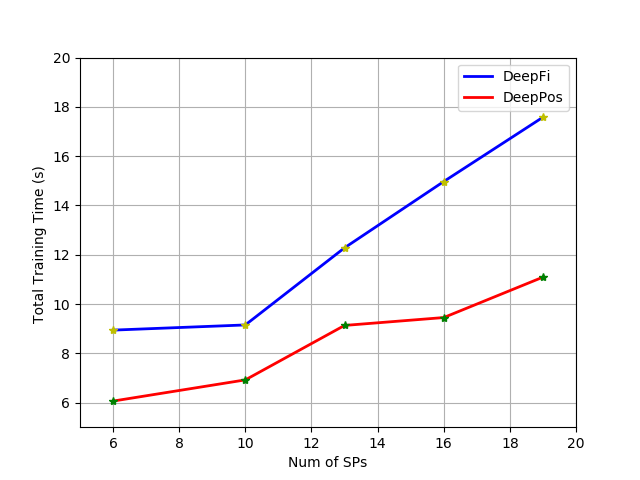}
		\caption*{a) Classroom}
	\end{minipage}
	\hfill
	\begin{minipage}[b]{0.23\textwidth}
		\includegraphics[width=\textwidth]{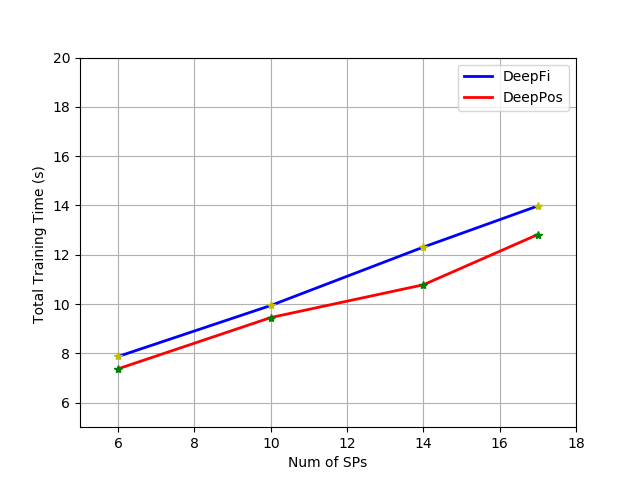}
		\caption*{b) Hall and Corridor}
	\end{minipage}
	\caption{Training stage time spent for parallel test}
	\label{time-test2}
\end{figure}

\begin{figure}[!tbp]
	\centering
	\begin{minipage}[b]{0.23\textwidth}
		\includegraphics[width=\textwidth]{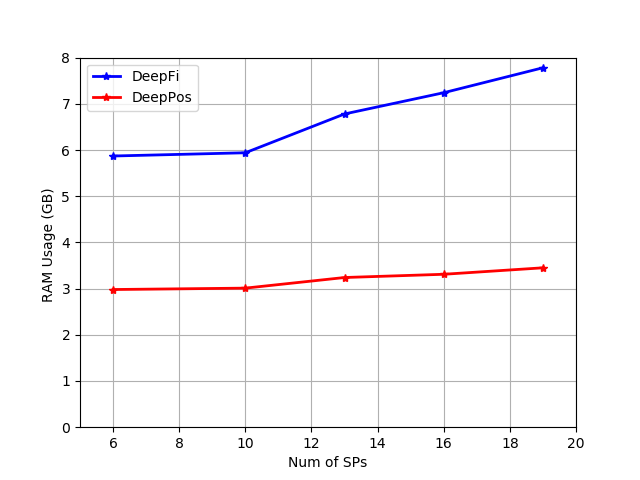}
		\caption*{a) Classroom}
	\end{minipage}
	\hfill
	\begin{minipage}[b]{0.23\textwidth}
		\includegraphics[width=\textwidth]{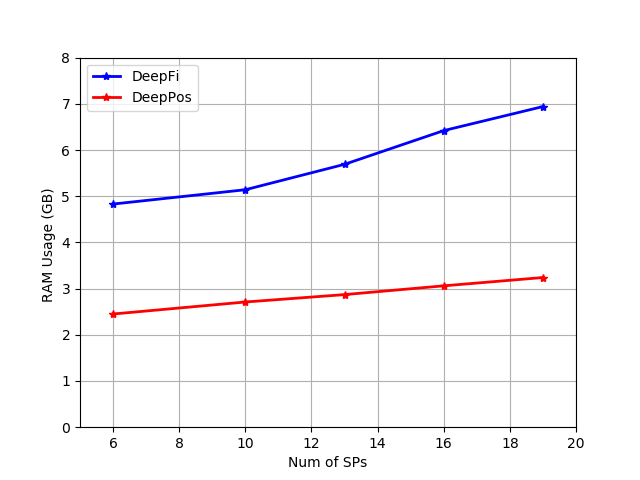}
		\caption*{b) Hall and Corridor}
	\end{minipage}
	\caption{Training stage memory usage  for parallel test}
	\label{ram-test2}
\end{figure}

Another possible assumption for DeepFi is to train SPs' models in parallel. The average training time and memory usage for this case is depicted in figures \ref{time-test2} and \ref{ram-test2}. As expected, in this case the training time of both schemes are getting closer together; however, the memory usage of DeepPos is much better than DeepFi (due to the need for training of one versus several models). Note that the extra training time of the DeepFi is probably because it uses more memory, and due to memory limitation of our computer (8GB), it gets some penalty due to virtual memory speed and heavier CPU utilization; Otherwise, DeepFi should have similar training time as DeepPos and the superiority of DeepPos in this mode is mainly on lower memory usage during training stage.   \\


\subsubsection{\textbf{Evaluating Testing Overhead}}
Increasing the number of SPs can boost the accuracy of the system. The point to consider though is by the extra training SPs, the model should not face much timing overhead, specially in testing stage. 

To study this behavior for both DeepFi and DeepPos systems, we conducted an experiment that we increased the number of SPs and for each case calculated the average (as mentioned in previous section) localization accuracy and also measure the execution time of the testing phase. 

\begin{figure}[!tbp]
	\centering
	\begin{minipage}[b]{0.23\textwidth}
		\includegraphics[width=\textwidth]{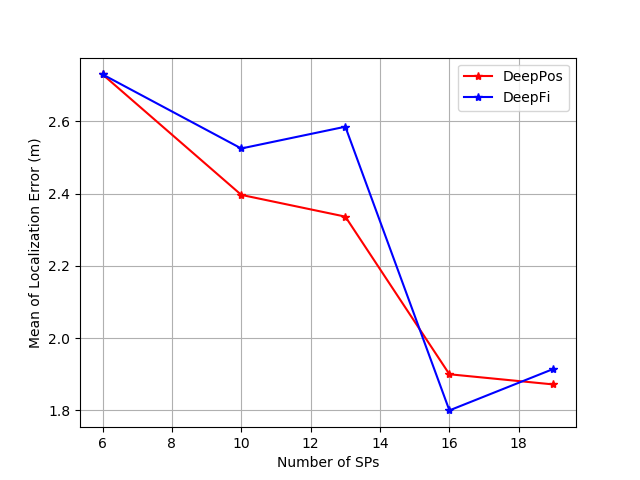}
		\caption*{ a) Classroom}
	\end{minipage}
	\hfill
	\begin{minipage}[b]{0.23\textwidth}
		\includegraphics[width=\textwidth]{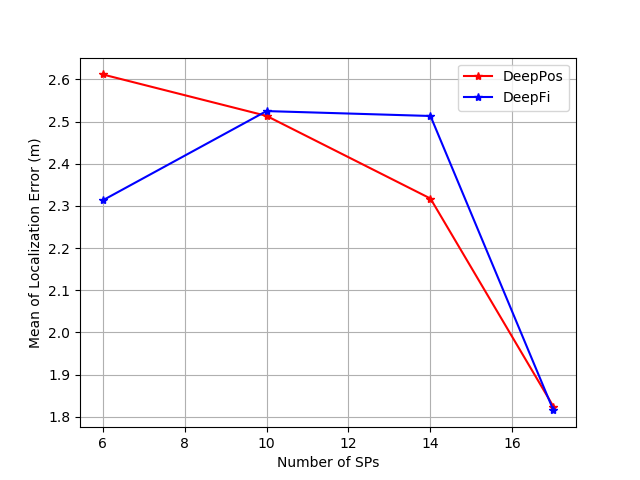}
		\caption*{ b) Hall and Corridor}	
	\end{minipage}
	\caption{Mean of localization error}
	\label{tb-err}
\end{figure}

\begin{figure}[!tbp]
	\centering
	\begin{minipage}[b]{0.23\textwidth}
		\includegraphics[width=\textwidth]{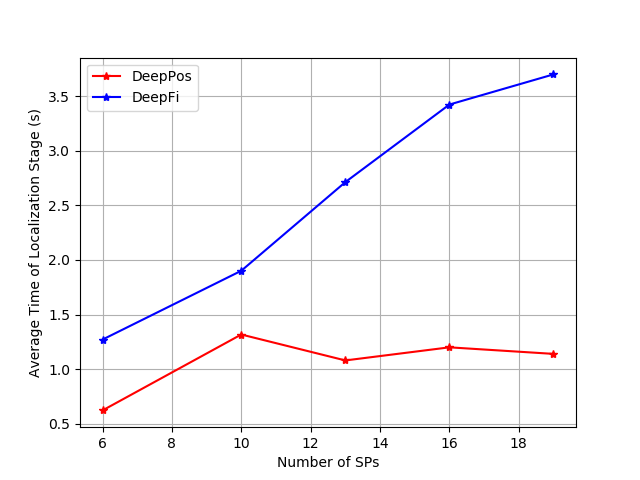}
		\caption*{ a) Classroom}
	\end{minipage}
	\hfill
	\begin{minipage}[b]{0.23\textwidth}
		\includegraphics[width=\textwidth]{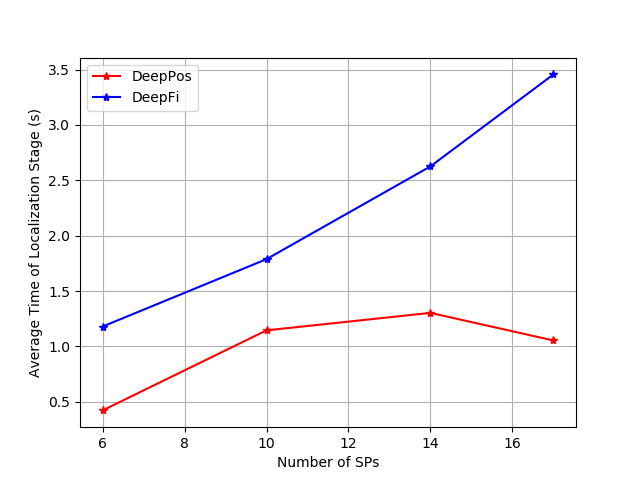}
		\caption*{ b) Hall and Corridor}	
	\end{minipage}
	\caption{Average time of localization stage}
		\label{tb-t-err}
\end{figure}

The results are depicted in figures \ref{tb-err} and \ref{tb-t-err}. Comparing the results, it can be verified that by addition of new SPs, the overall precision of both schemes increases almost similarly; however, DeepPos outperforms DeepFi in terms of prediction time.

\section{Conclusion and future works}

In this paper, we proposed DeepPos, a deep supervised autoencoder based system for indoor localization. DeepPos uses 2.4GHz WiFi fingerprinting measurements for training the model, but unlike similar approaches in this method, fingerprints are not stored for each of the training points. Alternatively, we fit only one model to the environment that uses deep SAE structure (the encoding layers act as a feature extractor, user locations are labels which are appended to this extracted features, and decoding layers play the role of a supervised neural network). The coordinates of a test location is then determined by finding the labels of the SAE that results in lowest CSI regeneration error. The performance of the proposed scheme is evaluated in two indoor environments and we have shown it keeps the high accuracy while it has much less complexity compared to other deep learning based indoor positioning methods.




\ifCLASSOPTIONcaptionsoff
  \newpage
\fi



\bibliographystyle{IEEEtran}  
\bibliography{myreferences}




\end{document}